\documentclass[aps,prb,twocolumn,showpacs,groupedaddress,superscriptaddress]{revtex4}
\usepackage{graphics,bm}
\usepackage{amssymb}
\usepackage{amsmath}
\usepackage{epsfig}
\usepackage{epsf}
\usepackage[usenames]{color}

\begin{document}

\title{Spin motive forces due to magnetic vortices and domain walls}

\author{M.E. Lucassen}
\email{m.e.lucassen@uu.nl}

\affiliation{Institute for Theoretical Physics, Utrecht
University, Leuvenlaan 4, 3584 CE Utrecht, The Netherlands}

\author{G.C.F.L. Kruis}

\author{R. Lavrijsen}

\author{H.J.M. Swagten}

\author{B. Koopmans}

\affiliation{Department of Applied Physics, Center for NanoMaterials and COBRA Research Institute, Eindhoven University of Technology, P.O. Box 513, 5600 MB Eindhoven, The Netherlands}

\author{R.A. Duine}

\affiliation{Institute for Theoretical Physics, Utrecht
University, Leuvenlaan 4, 3584 CE Utrecht, The Netherlands}

\date{\today}

\begin{abstract}
We study spin motive forces, {\it i.e.,} spin-dependent forces, and voltages induced by time-dependent magnetization textures, for moving magnetic vortices and domain walls. First, we consider the voltage generated by a one-dimensional field-driven domain wall. Next, we perform detailed calculations on field-driven vortex domain walls. We find that the results for the voltage as a function of magnetic field differ between the one-dimensional and vortex domain wall. For the experimentally relevant case of a vortex domain wall, the dependence of voltage on field around Walker breakdown depends qualitatively on the ratio of the so-called $\beta$-parameter to the Gilbert damping constant, and thus provides a way to determine this ratio experimentally. We also consider vortices on a magnetic disk in the presence of an AC magnetic field. In this case, the phase difference between field and voltage on the edge is determined by the $\beta$ parameter, providing another experimental method to determine this quantity.
\end{abstract}


\maketitle

% definitions
\def\m{{\bf m}}
\def\s{{\rm s}}
\def\sd{{\rm sd}}
\def\F{{\bf F}}
\def\nabvec{{\boldsymbol{\nabla}}}
\def\P{{\mathcal P}}
\def\l{{\ell}}
\def\x{{\bf x}}
\def\ii{{\hat\i}}
\def\jj{{\hat\j}}
\def\B{{\bf B}}

\section{Introduction}

One of the recent developments in spintronics is the study of spin
motive forces\cite{barnes2007} and spin pumping\cite{tserkovnyak2002}. These effects lead to the generation of charge and spin currents due to time-dependent magnetization textures. The idea of spin motive forces due to domain walls
is easily understood on an intuitive level: if an applied current
induces domain-wall motion,\cite{berger1984,berger1985,slonczewski1996,berger1996,zhang2004,barnes2005} Onsager's reciprocity theorem tells us
that a moving domain wall will induce a current. This idea was
already put forward in the eighties by Berger\cite{berger1986}.
In the case of a domain wall driven by large magnetic fields ({\it i.e.,} well above the so-called Walker breakdown field), a fairly simple approach to the
problem is justified where one goes to a frame of reference in which the spin quantization axis follows the magnetization texture.\cite{stern1992} This transformation gives rise to a
vector potential from which effective electric and magnetic fields are derived. Experimentally, the domain-wall induced voltage has recently been measured above Walker breakdown\cite{yang2009}, and the results are consistent with this approach. It has also been shown that the induced voltage well above Walker breakdown is determined from a topological argument that follows from the properties of the above-mentioned vector potential.\cite{yang2010}

The above approach only captures the reactive contribution to the spin-motive forces. When the velocity of the domain wall is below or just above Walker
breakdown, a theory is needed that includes more contributions to the spin motive forces. Renewed interest
has shed light on the non-adiabatic and dissipative
contributions to the spin motive
forces\cite{saslow2007,duine2008a,tserkovnyak2008} that are important in this regime. In this paper, we study this regime.

The article is organized as follows. In Section~\ref{sec:model} we summarize earlier results that give a general framework to compute electrochemical potentials for given time-dependent magnetization textures. In Section~\ref{sec:rigid} we consider an analytical model for a one-dimensional domain wall and numerically determine the form of the spin accumulation and the electrochemical potential. The results agree with the known results for the potential difference induced by a moving one-dimensional domain wall.\cite{duine2008a} In Section~\ref{sec:vortex} we turn to two-dimensional systems and study a vortex domain wall in a permalloy strip. We use a micro-magnetic simulator to obtain the magnetization dynamics, and numerically evaluate the
reactive and dissipative contributions to the voltage below and just above Walker breakdown and compare with experiment.\cite{yang2009} Another example of a two-dimensional system is a vortex on a disk which we treat in Section~\ref{sec:disk}. For small enough disks, the magnetic configuration is a vortex. Both experimentally and theoretically, it has been shown that a vortex driven by an oscillating magnetic field will rotate around its equilibrium position.\cite{Choe2004,Waeyenberge2006,KiSukLee2007, Kruger2007,Bolte2008} This gives rise to a voltage difference between the disk edge and center as was recently discussed by Ohe {\it et al.} [\onlinecite{Ohe2009}]. Here, we extend this study by including both the reactive and the dissipative contributions to the voltage, that turn out to have a relative phase difference. This gives rise to a phase difference between the drive field and voltage that is determined by the so-called
$\beta$-parameter.

\section{Model}\label{sec:model}

The spin-motive force field $\F(\vec{x})$ induced by a time-dependent magnetization texture that is
characterized at position $\vec{x}$ by a unit-vector magnetization direction
$\m(\vec{x},t)$ is given by\cite{duine2008a,tserkovnyak2008}
\begin{align}\label{eq:forcefield}
F_i=\frac{\hbar}{2}\left[\m\cdot(\partial_t\m\times\nabla_i\m)+\beta(\partial_t
\m\cdot\nabla_i\m)\right]\;.
\end{align}
This force field acts in this form on the majority spins, and with opposite sign on minority spins.
In this expression, the first term is the well-known reactive
term.\cite{barnes2007} The second term describes dissipative
effects due to spin relaxation\cite{duine2008a,tserkovnyak2008} and is proportional to the phenomenological $\beta$-parameter, which plays an important role in current-induced domain-wall motion.\cite{berger1984,berger1985,slonczewski1996,berger1996,zhang2004,barnes2005}
The spin accumulation $\mu_{\s}$ in the system follows from\cite{tserkovnyak2008}
\begin{align}\label{eq:spinaccumulation}
\frac{1}{\lambda_\sd^2}\mu_\s-\nabla^2\mu_\s=-\nabvec\cdot\F\;,
\end{align}
where $\lambda_\sd=\sqrt{\tau D}$ is the spin-diffusion length,
with $\tau$ a characteristic spin-flip time and $D$ the effective
spin-diffusion constant. Here, we assume that the spin-relaxation
time is much smaller than the timescale for magnetization
dynamics. The total electrochemical potential $\mu$ that is generated by the
spin accumulation due to a non-zero current polarization in the
system is computed from\cite{tserkovnyak2008}
\begin{align}\label{eq:electrochemical}
-\nabla^2\mu=\P(\nabla^2\mu_\s-\nabvec\cdot\F)\;,
\end{align}
where the current polarization is given by
$\P=(\sigma_\uparrow-\sigma_\downarrow)/(\sigma_\uparrow+\sigma_\downarrow)$.
Note that there is no charge accumulation for $\sigma_\uparrow=\sigma_\downarrow$, with $\sigma_\uparrow$($\sigma_\downarrow$) the conductivity of the majority (minority) spin electrons.

The magnetization dynamics is found from the Landau-Lifschitz-Gilbert (LLG) equation given by
\begin{align}\label{eq:LLG}
\frac{\partial\m}{\partial t}=\m\times\left(-\frac{\partial E_{\rm mm}[\m]}{\hbar\partial\m}\right)-\alpha\m\times\frac{\partial\m}{\partial t}\;.
\end{align}
Here, $E_{\rm mm}[\m]$ is the micromagnetic energy functional that includes exchange interaction, anisotropy, and external field, and $\alpha$ is the Gilbert damping constant.

\section{One-dimensional domain wall}\label{sec:rigid}

For one-dimensional problems the voltage difference can be easily found. For example, an analytic expression for the electric current (which is the open-circuit equivalent of the chemical-potential difference) was obtained by one of us for an analytical model for a one-dimensional driven domain wall.\cite{duine2008a} In this section, we solve the potential problem for a one-dimensional domain wall and obtain the explicit position dependence of the spin accumulation and the chemical potential.

A one-dimensional domain wall
($\partial_y\m=\partial_z\m=0$) is described by\cite{tatara2004}
\begin{align}\label{eq:ansatz}
\theta(x,t)=2{\rm arctan}\left\{
e^{Q[x-X(t)]/\lambda}\right\}\;,\qquad\phi(x)=0\;,
\end{align}
with
$\mathbf{\m}=(\sin\theta\cos\phi,\sin\theta\sin\phi,\cos\theta)$. Here, $Q=\pm 1$ is called the topological charge of the domain wall since it indicates the way in which an external field affects the domain-wall motion, {\it i.e.}, a field in the direction $+\hat{z}$ will move a domain wall in the direction $Q\hat{x}$. Here, we choose $Q=1$. The domain wall width is indicated by $\lambda$.

To study the time-evolution of a domain-wall, we let $\phi(x)\rightarrow\phi_0(t)$ so that the wall is described by time-dependent collective coordinates $\{X(t),\phi_0(t)\}$, called position and chirality, respectively. For external fields smaller than the Walker-breakdown field, there is no domain-wall precession ({\it i.e.,} the chirality is constant), and the domain wall velocity $v$ is constant so that $\partial_t\m=-v\partial_x\m$. Since
$\partial_y\m=\partial_z\m=0$, we immediately see that the first term on the right-hand side of Eq.~({\ref{eq:forcefield}) vanishes, and that the force is pointing along the x-axis. We then find that
$F_x=(\beta v\hbar/2)/ (\lambda^2\cosh[x/\lambda]^2)$.
Due to symmetry we have that
$\partial_y\mu=\partial_z\mu =\partial_y\mu_\s=\partial_z\mu_\s=0$. In Figs.~\ref{fig:spinaccumulationrigid}~and~\ref{fig:potentialrigid}
we plot the spin accumulation and the electrochemical potential as a function of $x$.
\begin{figure}[h!]
\centering
\includegraphics[width=8.5 cm]{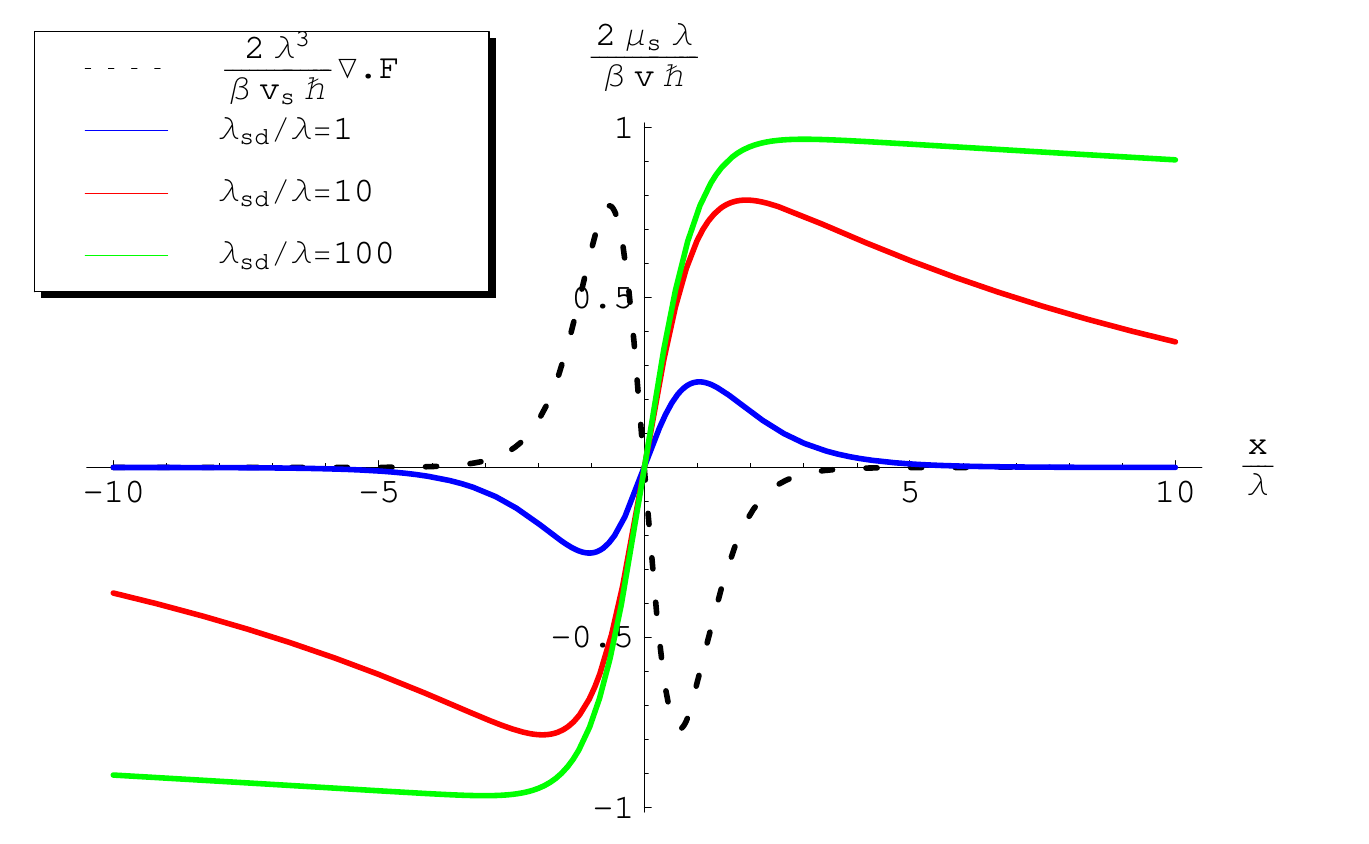}
\caption{(color online) Spin accumulation as a function of position for several values of the
spin-diffusion length. The black dotted line gives the value of
the source term. The spin accumulation tends to zero for
$x\rightarrow\pm\infty$.}\label{fig:spinaccumulationrigid}
\end{figure}
\begin{figure}[h!]
\centering
\includegraphics[width=8.5 cm]{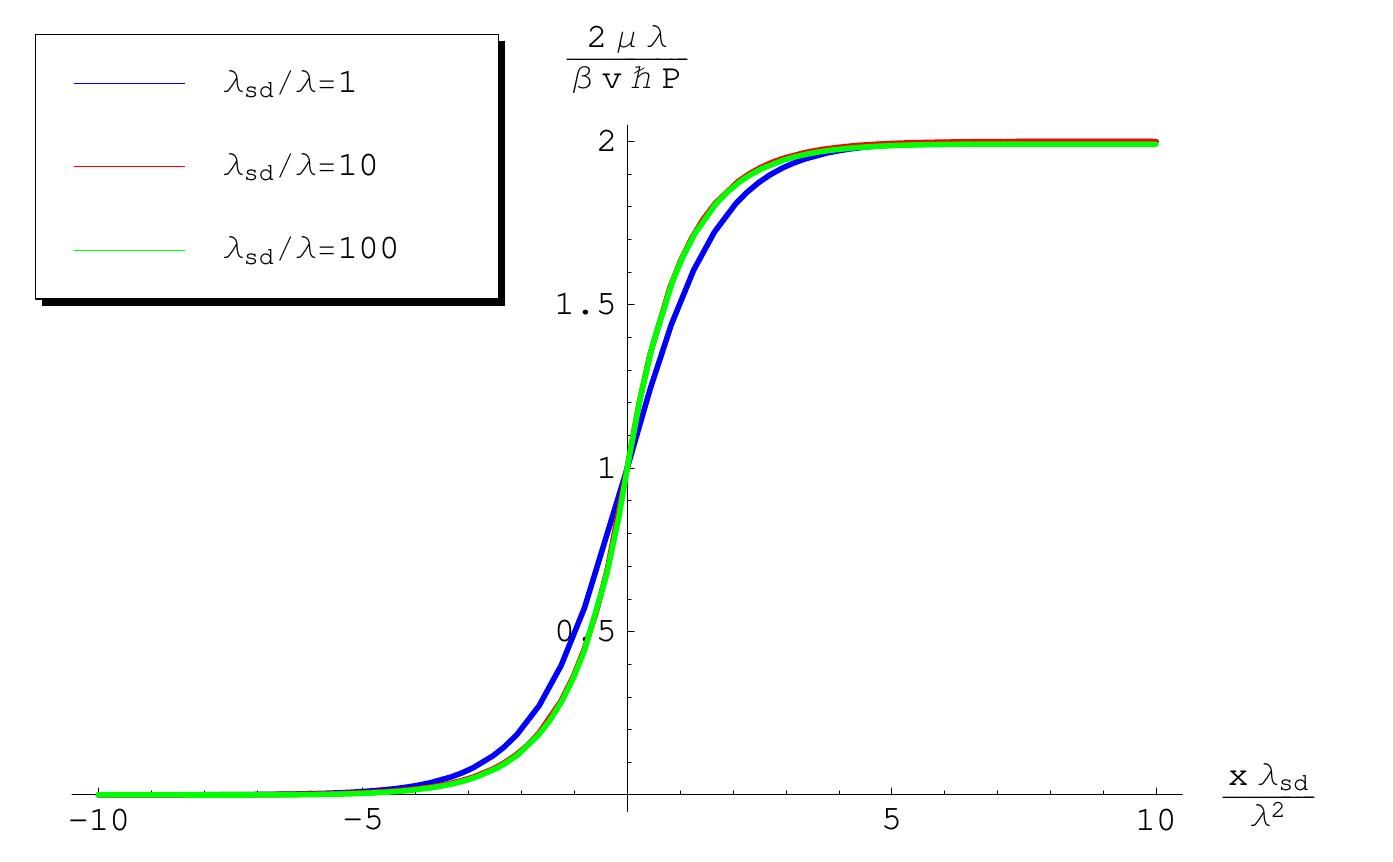}
\caption{(color online) Electrochemical potential as a function of position. Note that the potential is
proportional to the polarization and that on the horizontal axis the position $x$ is multiplied by the spin-diffusion length.}\label{fig:potentialrigid}
\end{figure}\\\\
From Fig.~\ref{fig:potentialrigid} we see that the total potential
difference
$\Delta\mu=\mu(x\rightarrow\infty)-\mu(x\rightarrow-\infty)$ is
independent of the spin-diffusion length and linear in the parameter $\beta$: $\Delta\mu=\hbar\P \beta v/\lambda$. Note that this result is only valid below Walker Breakdown.

To find the voltage for all fields $B$ we generalize the results for the voltage difference in Ref.~[\onlinecite{duine2008a}] for general domain-wall charge $Q$. A general expression for the voltage in one dimension is given by\cite{duine2008a}
\begin{align}\label{eq:deltaV}
\Delta \mu=-\frac{\hbar \P}{2|e|}\int dx\left[\m\cdot(\partial_t\m\times\partial_x\m)+\beta \partial_t\m\cdot\partial_x\m\right]\;.
\end{align}
We insert the {\it ansatz} [Eq.~(\ref{eq:ansatz}) with $\phi(x)=\phi_0(t)$] into Eq.~(\ref{eq:deltaV}) and find
\begin{align}
\Delta\mu=-\frac{\hbar \P}{2|e|}\left[Q\dot{\phi}_0+\beta \frac{\dot{X}}{\lambda}\right]\;.
\end{align}
To find a time-averaged value for the voltage we consider the equations of motion for a domain wall that is driven by a transverse magnetic field $B$\cite{walker1974,tatara2004,tatara2008}, contributing to the energy $-g\B\cdot\m$, with $g>0$. The equations of motion for $X(t)$ and $\phi_0(t)$ are ultimately derived from the LLG equation in Eq.~(\ref{eq:LLG}), and given by
\begin{align}
(1+\alpha^2)\dot{\phi}_0&= -\frac{gB}{\hbar}-\alpha\frac{K_\perp}{2\hbar}\sin (2\phi_0)\;,\nonumber\\
(1+\alpha^2)\frac{\dot{X}}{\lambda}&=\alpha Q\frac{gB}{\hbar}-Q\frac{K_\perp}{2\hbar}\sin (2\phi_0)\;.
\end{align}
Here, $K_\perp$ is the out-of-plane anisotropy constant. These equations are solved by
\begin{align}
\langle\dot{\phi}_0\rangle&=-\frac{{\rm Sign}(B)}{1+\alpha^2}{\rm Re}\left[\sqrt{\left(\frac{gB}{\hbar}\right)^2-\left(\frac{\alpha K_\perp}{2\hbar}\right)^2}\right]\;,\nonumber\\
&\langle\dot{X}\rangle=\frac{\lambda Q}{1+\alpha^2} \left(\frac{gB}{\alpha\hbar} +\frac{\langle\dot{\phi}_0\rangle}{\alpha}\right)
\end{align}
where $\langle..\rangle$ denotes a time average. It follows that the voltage difference for general topological charge is
\begin{align}
\Delta \mu&=-{\rm Sign}(B)\frac{Q}{1+\alpha^2}\frac{\hbar \P}{2|e|}\Bigg\{\frac{\beta}{\alpha}\frac{g|B|}{\hbar}\nonumber\\
&-\left(1+\frac{\beta}{\alpha}\right){\rm Re}\left[\sqrt{\left(\frac{gB}{\hbar}\right)^2-\left(\frac{\alpha K_\perp}{2\hbar}\right)^2}\right]\Bigg\}\;.
\end{align}
Note that the overall prefactor Sign$(B)Q$ makes sense: inversion of the magnetic field should have the same result as inversion of the topological charge.

In the above, we used a domain-wall ansatz with magnetization perpendicular to the wire direction. Using the topological argument by Yang et al.\cite{yang2010} one can show that the result is more general and holds also for head-to-head and tail-to-tail domain walls. Therefore, for a one-dimensional domain wall, the reactive and dissipative contributions, {\it i.e.,} the contributions with and without $\beta$ in the above expression, to the voltage always have opposite sign.

\section{Vortex domain wall}\label{sec:vortex}
For more complicated two-dimensional structures the spin-motive force field can have rotation and the simplified expression in Eq.~(\ref{eq:deltaV}) is no longer valid so that we need to treat the full potential problem in Eqs.~(\ref{eq:forcefield}-\ref{eq:electrochemical}). Motivated by recent experimental results\cite{yang2009} we consider in this section the voltage induced by a moving vortex domain wall.

We study the magnetization dynamics using a micro-magnetic simulator\cite{scheinfein} from which we obtain the magnetization $\m(\vec{x},t)$. This simulator solves the LLG equation in Eq.~(\ref{eq:LLG}). For comparison with the experiment by Yang {\it et al}.,\cite{yang2009} we simulate a permalloy sample that has the same dimensions as this experiment, i.e. 20nm $\times$ 500nm $\times$ 32$\mu$m, which is divided in 1 $\times$ 128 $\times$ 8192 lattice points. On this sample, we drive a head-to-head vortex domain wall by means of a magnetic field that is pointing from right to left, such that the vortex moves from right to left. For several field strengths, we obtain the magnetization $\m$, and its time-derivative which allows us to compute the force field $\F$ at each lattice point. Next, we solve the matrix problem that is the discrete equivalent to the potential problem in Eqs.~(\ref{eq:spinaccumulation})~and~(\ref{eq:electrochemical}). For details on this calculation, see App.~\ref{app:lattice}.

We first investigate the velocity of the vortex domain wall as a function of the applied field. We use the value $\alpha=0.02$ for the Gilbert-damping parameter to obtain the curve in Fig.~\ref{fig:velocity}.
\begin{figure}[h!]
\centering
\includegraphics[width=8.5 cm]{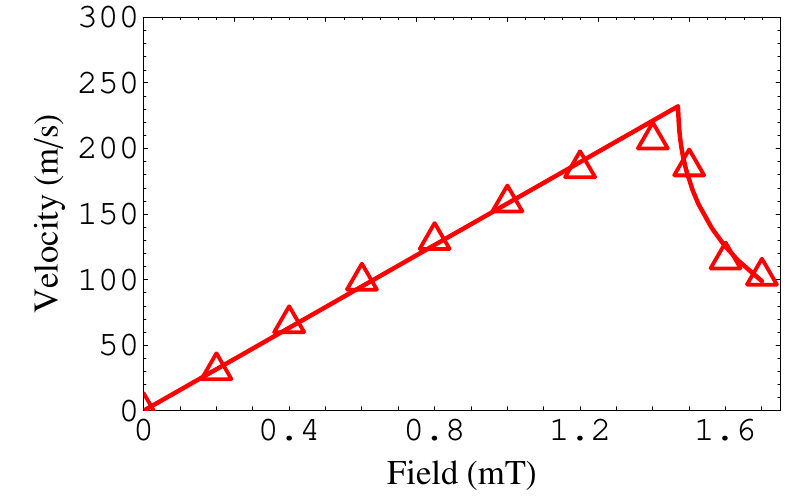}
\caption{Velocity of the vortex domain wall as a function of the magnetic field strength for $\alpha=0.02$. Above Walker breakdown, the velocity is time-averaged. The line is a guide to the eye.}\label{fig:velocity}
\end{figure}\\\\
The decrease in velocity for $B=1.5$ mT signals Walker breakdown. Indeed, up to fields $B=1.4$ mT, the vortex moves parallel to the long direction of the sample. For $B=1.5$ mT, the vortex domain wall motion is more complicate and has a perpendicular component.\cite{beach2005,clarke2008} We therefore expect that below Walker Breakdown, just like for the one-dimensional domain wall, the vortex domain wall only has a dissipative contribution to the voltage. Comparison with the experimental results of Ref.~[\onlinecite{yang2009}] shows that our velocity is roughly a factor 2 higher. This might be partly caused by a difference in damping and partly by the presence of defects in the experiment which causes pinning and therefore a decrease of velocity. The exact value of the Walker breakdown field is hard to compare, since this depends also on the exact value of the anisotropy. Nonetheless our value for the Walker breakdown field is of the same order as Ref.~[\onlinecite{yang2009}]. Moreover, what is more important is the dependence of wall velocity and wall-induced voltage on the magnetic field normalized to the Walker-breakdown field, as these results depend less on system details.

\begin{figure}[h!]
\centering
\includegraphics[width=8.5 cm]{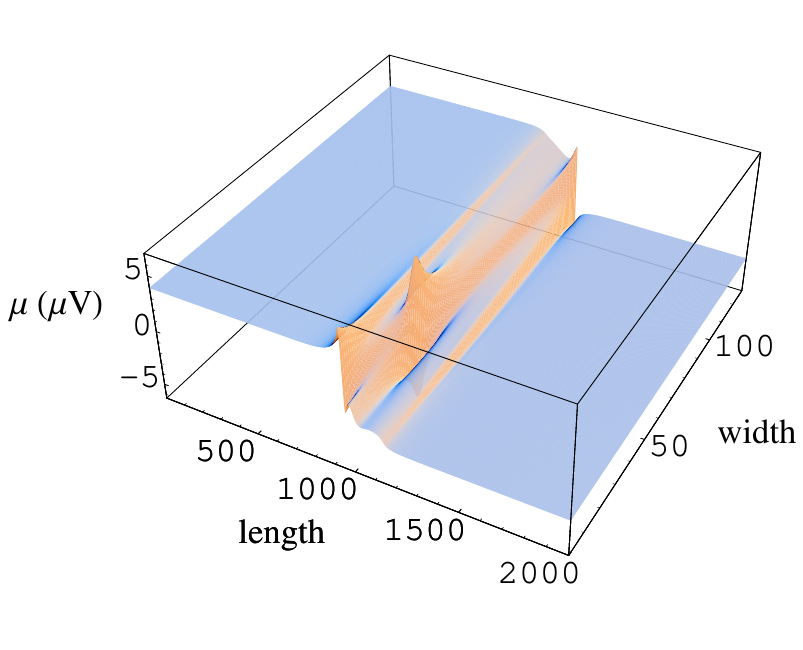}
\caption{Electrochemical potential as a function of position for a moving vortex domain wall on the sample. The numbers on the horizontal axes correspond to lattice points with separation $a=3.9$nm. This specific figure is for $\alpha=0.02$, $H=0.8$mT (i.e. below Walker breakdown), $\mathcal{P}=1$ and $\lambda_{\rm sd}=a$. Note that the peak signals the position of the vortex core.}\label{fig:mooieplaatjes}
\end{figure}
An example of a specific form of the electrochemical potential on the sample due to a field-driven vortex domain wall is depicted in Fig.~\ref{fig:mooieplaatjes}. We see that there is a clear voltage drop along the sample, like in the one-dimensional model. Additionally, the potential shows large gradients around the vortex core and varies along the transverse direction of the sample. For each field strength, we compute the voltage difference as a function of time. For field strengths below Walker Breakdown we find that, as expected, only the dissipative term contributes and the voltage difference rapidly approaches a constant value in time. This is understood from the fact that in this regime, the wall velocity is constant after a short time. The dissipative contribution to the voltage is closely related to the velocity along the sample, as can be seen in Fig.~\ref{fig:potentialtime}.
\begin{figure}[h!]
\centering
\includegraphics[width=8.5 cm]{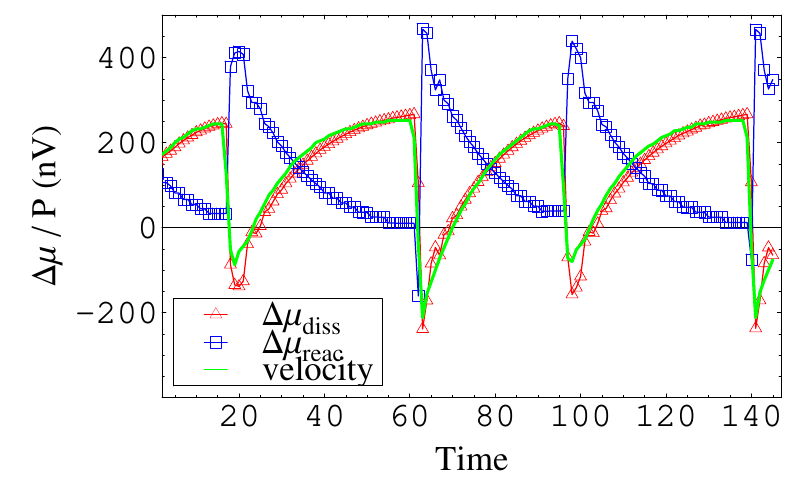}
\caption{(color online) reactive (blue squares) and dissipative (red triangles) contributions to the voltage as a function of time. The numbers on the horizontal axis correspond to time steps of 0.565 ns. The green line gives the velocity along the sample, it is scaled to show the correlation with the voltage. These curves are taken for $\alpha=0.02$, $\beta=\alpha$ and field strength $B=1.6$mT.}\label{fig:potentialtime}
\end{figure}

Above Walker breakdown the reactive term contributes. We find that for $\beta=\alpha$, the oscillations in the reactive component compensate for the oscillations in the dissipative component. If we look closely to Fig.~\ref{fig:potentialtime} we see that the length of the periods is not exactly equal. The periods correspond to a vortex moving to the upper edge of the sample, or to the lower edge. The difference is due to the initial conditions of our simulation.
We average the voltage difference over time to arrive at the result in Fig.~\ref{fig:potential}.
\begin{figure}[h!]
\centering
\includegraphics[width=8.5 cm]{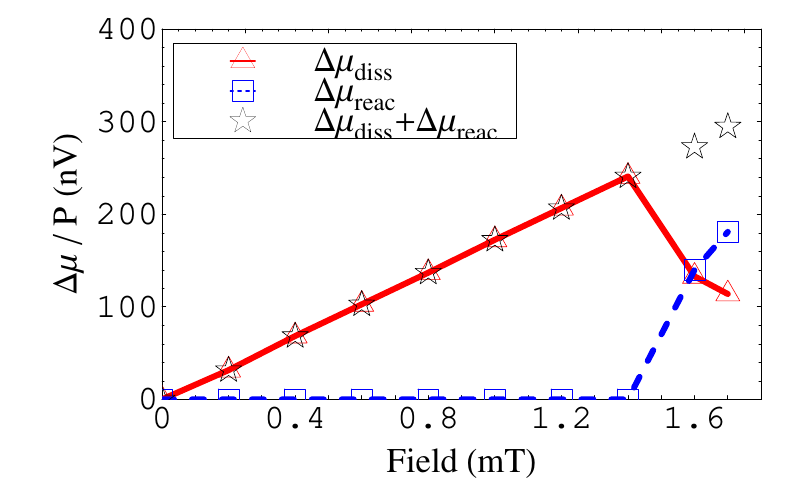}
\caption{(color online) Voltage drop along the sample for $\alpha=0.02$ and $\beta=\alpha$.}\label{fig:potential}
\end{figure}
We see that the dissipative contribution becomes smaller for fields larger than the Walker breakdown field, whereas the reactive contribution has the same sign and increases. In fact, for $\beta=\alpha$, the reduction of the dissipative contribution is exactly compensated by the reactive contribution. The $\beta$ dependence is illustrated in Fig.~\ref{fig:potentialfuncbeta}.
\begin{figure}[h!]
\centering
\includegraphics[width=8.5 cm]{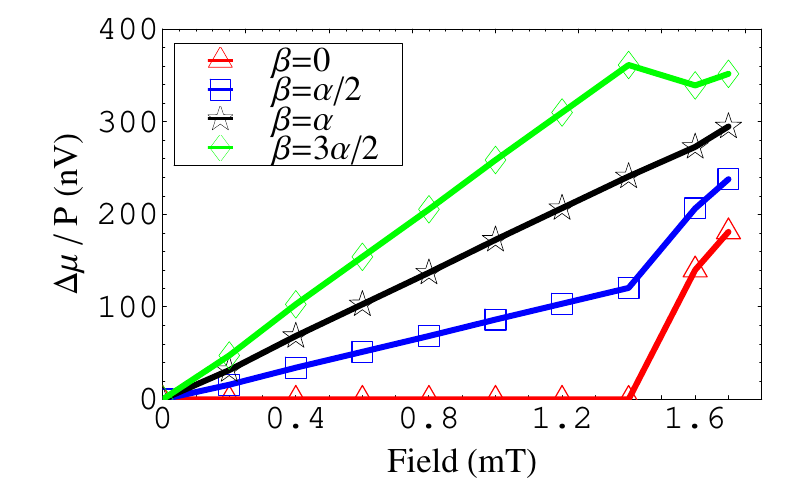}
\caption{(color online) Voltage drop along the sample for $\alpha=0.02$ and several values for $\beta$.}\label{fig:potentialfuncbeta}
\end{figure}
The behavior is fundamentally different from the one-dimensional domain-wall situation: for the vortex domain wall, the dissipative contribution has the same sign as the reactive contribution.

In order to understand the relative sign, we now discuss general vortex domain walls. A single vortex (i.e. with vorticity $q=+1$) is described by two parameters: the charge $p=\pm 1$ indicates wether the central magnetic moment points in the positive or negative $z$ direction and the chirality $C$ indicates wether the magnetic moments align in a clockwise ($C=-1$) or anti-clockwise ($C=+1$) fashion. We have a vortex that is oriented clockwise $C=-1$. The relative sign is explained from a naive computation of the voltage above Walker breakdown that does not take into account the rotation of the spin-motive force field
\begin{align}
\Delta \mu&\propto-\int d\x\left[\beta \partial_t\m\cdot\partial_x\m+\m\cdot(\partial_t\m\times\partial_x\m) \right]\nonumber\\
&=\beta v_x \int d\x (\partial_x\m)^2+v_y\int d\x\m\cdot(\partial_y\m\times\partial_x\m)\nonumber\\
&\propto (\beta\delta + 1) v_x \;.
\end{align}
where $\delta$ is a positive number and we used that above Walker breakdown $\m\simeq\m(x-v_xt,y-v_yt)$ with $v_x\neq 0$ and $v_y\neq 0$. Note that if $v_y=0$ (below Walker breakdown) the reactive term, {\it i.e.}, the second term in the above expression, indeed vanishes. We used in the last line that above Walker breakdown $v_x\propto -pv_y$ and $\int d\x\m\cdot(\partial_y\m\times\partial_x\m)\propto -p$. The former equality is understood from a geometric consideration: consider a sample with a vortex characterized by $C=1$, $p=1$ and $v_xv_y<0$. By symmetry, this is equivalent to $C=-1$, $p=-1$ and $v_xv_y>0$. It is therefore clear that the sign of $v_xv_y$ depends on either the polarization, or the handedness of the vortex. Since we know from the vortex domain wall dynamics that reversal of the polarization reverses the perpendicular velocity,\cite{yang2009} we conclude that $v_xv_y$ does not depend on the handedness of the vortex. The latter equality is understood from a similar argument: $\iint dx dy\m\cdot(\partial_y\m\times\partial_x\m)$ changes sign under the transformation $\m\rightarrow-\m$. During this transformation, both $p\rightarrow-p$ and $C\rightarrow-C$, and therefore their product cannot account for the total sign reversal. Therefore, the integral depends on the polarization\cite{yang2010} but not on the handedness of the vortex. The positive number $\delta$ is obtained from our numerical simulation, which suggests that the magnetic-field dependence of the voltage is
\begin{align}
\Delta \mu= \beta B \times {\rm constant} +(1-\beta/\alpha)
|\Delta \mu_{\rm reactive}|\;.
\end{align}
Note that the sign of the relative contributions can also be obtained using the topological argument by Yang et al.\cite{yang2010}, which gives the same result.

We compare our results in Fig.~\ref{fig:potentialfuncbeta} with the experiment by Yang et al.\cite{yang2009}. If we assume that the voltage below Walker breakdown lies roughly on the same line as the voltages above Walker breakdown, their results suggest a slope of $10 $nV/Oe. For $\mathcal{P}\sim 0.8$, our results suggest a slope of $(\beta/\alpha)14 $nV/Oe. Taking into account our higher velocity, we find that $\beta$ in the experiment is somewhat larger than $\alpha$. The decrease in slope of the voltage in Ref.~[\onlinecite{yang2009}] as Walker breakdown is approached from above also suggests $\beta>\alpha$.

In conclusion, the behavior of the voltage around Walker breakdown allows us to determine the ratio $\beta/\alpha$. In experiment, the potential difference as a function of the applied magnetic field would show an upturn or downturn around Walker breakdown as in Fig.~\ref{fig:potentialfuncbeta}, which corresponds to $\beta<\alpha$ and $\beta>\alpha$, respectively.

\section{Magnetic vortex on a disk}\label{sec:disk}
On small disks (of size $\mu$m and smaller) of ferromagnetic material the lowest energy configuration is a vortex. It has been shown that one can let the vortex rotate around its equilibrium position by applying an AC magnetic field\cite{Choe2004,Waeyenberge2006,KiSukLee2007,Bolte2008,Kruger2007}. This motion gives rise via Eq.~(\ref{eq:forcefield}) to a spin motive force on the spins, which induces a voltage on the edge of the disk relative to a fixed reference voltage, {\it e.g.} the disk center. Ohe {\it et al.}\cite{Ohe2009} have shown that the reactive contribution to the spin motive force field can be seen as a dipole that is pointing in the radial direction, {\it i.e.,} the divergence of the force field consists of a positive and a negative peak along the radial direction (note that the divergence of the force field can be seen as an effective charge). Rotation of this dipole gives rise to an oscillating voltage on the edge of the sample. Here, we consider also the dissipative contribution to the voltage.

We consider a vortex on a disk with radius $R$ that moves around its equilibrium position ({\it i.e.,} the center of the disk) at a distance $r_0$ from the center of the disk with frequency $\omega$. We use as a boundary condition that the magnetization on the edge of the disk is pointing perpendicular to the radial direction. In equilibrium, the micro-magnetic energy density of the form $-J\m\cdot\nabla^2\m-K_\perp m_z^2$ is minimized by
\begin{align}\label{eq:magnetizationdisk}
m_x(x,y)&=
\frac{-y}{\sqrt{x^2+y^2}}\cos\left[2\arctan\left(e^{-C\sqrt{x^2+y^2}/\kappa}\right)\right] \nonumber\\
m_y(x,y)&=\frac{x}{\sqrt{x^2+y^2}}\cos\left[2\arctan\left(e^{-C\sqrt{x^2+y^2}/\kappa}\right)\right] \nonumber\\
m_z(x,y)&=p\sin\left[2\arctan\left(e^{\sqrt{x^2+y^2}/\kappa}\right)\right]
\;,
\end{align}
where the center of the vortex is chosen at $x=y=0$. Here $\kappa=\sqrt{K_\perp/J}$ is the typical width of the vortex core. For permalloy this length scale is of the order $\sim 10$nm. The parameters $p$ and $C$ are defined as before, for definiteness we choose $p=1$, $C=-1$. To describe clockwise circular motion of the vortex around its equilibrium position at fixed radius $r_0$ we substitute $x\rightarrow x-r_0\sin(\omega t)$ and $y\rightarrow y-r_0\cos(\omega t)$. Note that we assume that the form of the vortex is not changed by the motion, which is a good approximation for $r_0\ll R$.

From the magnetization in Eq.~(\ref{eq:magnetizationdisk}), we compute the force field using Eq.~(\ref{eq:forcefield}). The reactive and dissipative contributions to the divergence of the force field are shown in Fig.~\ref{fig:dipoles}.
\begin{figure}[h!]
\centering
\includegraphics[width=8.5 cm]{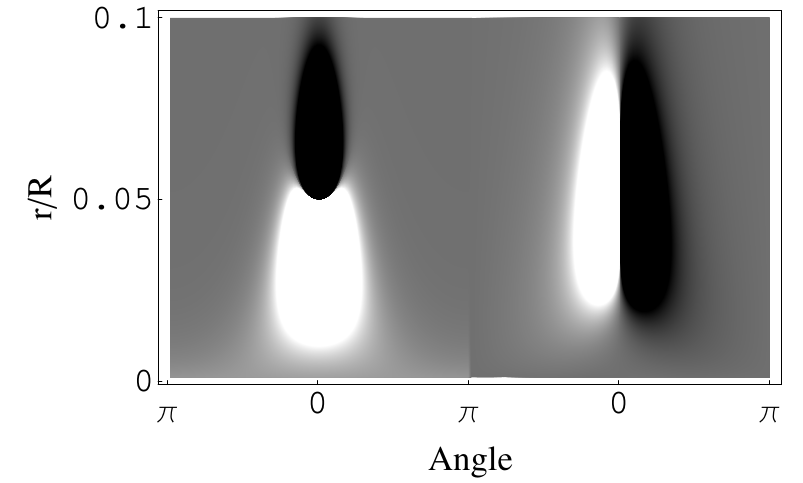}\\
\caption{The reactive (left) and dissipative (right) contributions to the divergence of the force field. White means positive values, black is negative values. The reactive contribution can be seen as a dipole in the radial direction. The dissipative contribution is a dipole perpendicular to the radial direction.}\label{fig:dipoles}
\end{figure}
The direction of the dipoles follows directly from Eq.~(\ref{eq:forcefield}) if we realize that for our system $-\partial_t\m\cdot\vec{\nabla}\m=\vec{v}(\partial_{\hat{v}}\m)^2$ is always pointing in the direction of the velocity which shows that the dissipative contribution points along the velocity. Likewise the reactive contribution is always pointing perpendicular to the velocity.

From the relative orientations of the effective dipoles, we expect that the peaks in the reactive and dissipative contributions to the voltage on the edge
will differ by a phase of approximately $\pi/2$ (for $r_0/R\rightarrow 0$ this is exact). We divide our sample in 1000 rings and 100 angles and use the general method in App.~\ref{app:lattice} to find the voltage on the edge shown in Fig.~\ref{fig:voltagedisk}.
\begin{figure}[h!]
\centering
\includegraphics[width=8.5 cm]{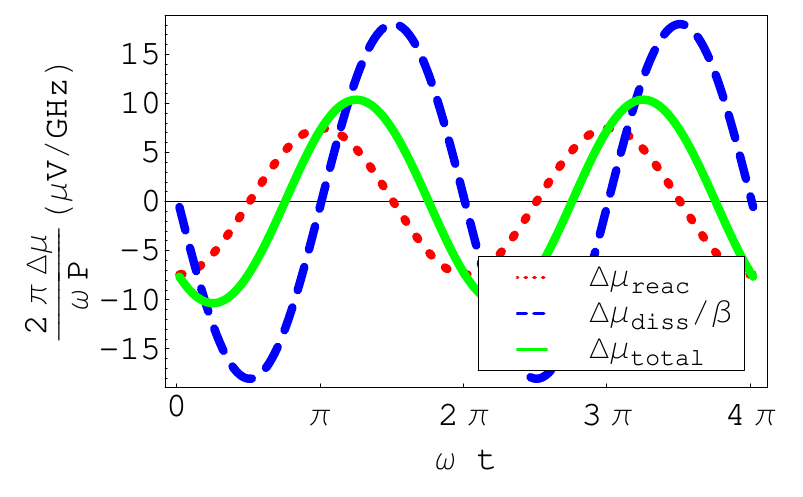}\\
\caption{(color online) The reactive (red dashed curve) and dissipative (blue dashed curve) contributions to the voltage difference between opposite points on the edge of the disk. The green line gives the total voltage difference $\Delta\mu_{\rm total}=\Delta\mu_{\rm reac}+\Delta\mu_{\rm diss}$, in this example for $\beta=0.4$. We used $r_0=10 \lambda_{\rm sd}$, $R=100 \lambda_{\rm sd}$ and $\kappa=\lambda_{\rm sd}$. For realistic spin-diffusion length $\lambda_{\rm sd}\simeq 5$ nm, these parameter values agree with the system of Ohe {\it et al.}\cite{Ohe2009}.}\label{fig:voltagedisk}
\end{figure}
To compare with Ref.~[\onlinecite{Ohe2009}], we take a frequency $\omega/(2\pi)=300$ MHz and $\mathcal{P}=0.8$, which yields amplitudes for the reactive contribution of $\sim \mu$V on the edge. However, Ohe {\it et al}. suggest that voltage probes that are being placed closer to the vortex core measure a higher voltage. This indeed increases the voltage up to order $\sim 10\;\mu$V at $r=2r_0$. Placing the leads much closer to the vortex core does not seem to be realistic because of the size of the vortex. Since the voltage scales with velocity, it can also be increased by a larger radius of rotation, i.e. by applying larger magnetic fields. However, for disks larger than $1\;\mu$m, the vortex structure is lost.

The dissipative contribution becomes important for large values of $\beta$. In principle, it is possible to determine $\beta$ by looking at the shift of the peak in the total voltage with respect to the peak in the reactive contribution, which is in turn determined by the phase of the applied magnetic field. The phase difference between applied field and measured voltege then behaves as $\tan(\Delta\phi)\propto\beta$.

\section{Discussion and Conclusion}\label{sec:discussion}

We have investigated the voltage that is induced by a field-driven vortex domain wall in detail. In contrast to a one-dimensional model of a domain wall, the reactive and dissipative contribution to the voltage have the same sign. The qualitative differences for different values of $\beta$ provide a way to determine the ratio $\beta/\alpha$ experimentally by measuring the wall-induced voltage as a function of magnetic field. To this end the experimental results in Ref.~[\onlinecite{yang2009}] are in the near future hopefully extended to fields below Walker breakdown, which is challenging as the voltages become smaller with smaller field.

We also studied a magnetic vortex on a disk. When the vortex undergoes a circular motion, a voltage is induced in the sample. Earlier work computed the reactive voltage on the edge of the disk,\cite{Ohe2009} here we include also the dissipative contribution to the voltage. We find that the phase difference between voltage and AC driving field is determined by the $\beta$-parameter.

\begin{acknowledgments}
This work was supported by the Netherlands Organization for
Scientific Research (NWO) and by the European Research Council (ERC) under the Seventh Framework Program (FP7).
\end{acknowledgments}

\appendix
\section{Boundary conditions}\label{app:boundary}
As a boundary condition for the potential problems, we demand that the total spin current and charge current
perpendicular to the upper and lower boundaries is zero: $j_{\rm s}^\perp=j_{\uparrow}^\perp-j_\downarrow^\perp=0$ and $j_{\rm s}^\perp=j_{\uparrow}^\perp+j_\downarrow^\perp=0$. Therefore, the majority-and minority spin currents are necessarily zero. They are given by $j_\uparrow=\sigma_\uparrow(F-\nabla\mu_\uparrow)$ and $j_\downarrow=\sigma_\downarrow(-F-\nabla\mu_\downarrow)$. From this, the boundary conditions on the derivatives of the potentials follow as $\partial_\perp\mu_s=\partial_\perp(\mu_\uparrow-\mu_\downarrow)/2=F$ and $\partial_\perp\mu=\partial_\perp(\mu_\uparrow+\mu_\downarrow)/2=0$.
We consider a two-dimensional sample that is infinitely long in the x-direction, and of finite size $2\Lambda$ in the y-direction the boundary conditions are
\begin{align}\label{eq:boundary}
\partial_y\mu_\s(x,y=\pm\Lambda)&=F_y(x,y=\pm\Lambda)\;.
\end{align}
To measure the induced voltage, we also put
the derivatives of the potential at infinity to zero so that the boundary conditions for the electrochemical potential are
\begin{align}\label{eq:boundarymu}
&\partial_y\mu(x,y=\pm\Lambda)=0\;,\nonumber\\
&\partial_x\mu(x\rightarrow\pm\infty,y)=0\;.
\end{align}

\section{Potential problem on a Lattice}\label{app:lattice}

We consider a two-dimensional lattice, where we have spin accumulation $\mu_{\rm s}^{i,j}$ and an electrochemical potential $\mu^{i,j}$ at site $i,j$. Between sites $(i,j)$ and $(i,j+1)$, there can be a particle current density of majority spins
\begin{align}
j_{\uparrow,\jj}^{i,j+1/2}&=\sigma_\uparrow\left(F_\jj^{i,j+1/2} +\frac{\mu_\uparrow^{i,j}-\mu_\uparrow^{i,j+1}} {a_\jj^{i,j+1/2}}\right)\nonumber\\
&=\sigma_\uparrow\left(F_\jj^{i,j+1/2} -\delta_\jj\mu_\uparrow^{i,j+1/2}\right)\;,
\end{align}
with $a_{\jj}^{i,j+1/2}$ the lattice spacing in the $\jj$ direction between sites $(i,j)$ and $(i,j+1)$, and a particle current density of minority spins
\begin{align}
j_{\downarrow,\jj}^{i,j+1/2}&=\sigma_\downarrow \left(-F_\jj^{i,j+1/2}+
\frac{\mu_\downarrow^{i,j} -\mu_\downarrow^{i,j+1}}{a_\jj^{i,j+1/2}}\right)\nonumber\\
&=\sigma_\downarrow\left(-F_\jj^{i,j+1/2} -\delta_\jj\mu_\uparrow^{i,j+1/2}\right)\;,
\end{align}
and equivalently for currents in the $\ii$ direction. The derivative $\delta_\jj$ is defined as $\delta_\jj O^{i,j}=(O^{i,j+1/2}-O^{i,j-1/2})/a_{\jj}^{i,j}$, and likewise for $\delta_\ii$. Note that upper indices $(i,j)$ denote a position on the lattices and lower indices $\ii$ or $\jj$ denote a direction. We can write $\mu_\uparrow=\mu+\mu_{\rm s}$ and $\mu_\downarrow=\mu-\mu_{\rm s}$. The continuity-like equations for the density of majority- and minority spins are (note that spins move in the direction of the current) \begin{align}
A^{i,j}\frac{n_{\uparrow\downarrow}^{i,j}}
{\tau}=-\frac{\Delta (\l^{i,j}j_{\uparrow\downarrow}^{i,j})}{|e|}\;,
\end{align}
with characteristic spin-flip time $\tau$ and with the dimensionless operator $\Delta$ given by
\begin{align}
\Delta O^{i,j}&=O^{i+1/2,j}_\ii-O^{i-1/2,j}_\ii +O^{i,j+1/2}_\jj-O^{i,j-1/2}_\jj\;.
\end{align}
These definitions allows for non-square lattices with sides at position $(i\pm 1/2,j)$ or $(i,j\pm 1/2)$ that have length $\l_\ii^{i\pm 1/2,j}$ or $\l_\jj^{i,j\pm 1/2}$ (lower index denotes the normal direction), respectively, and the area of the site itself given by $A^{i,j}$.

The equation for the electrochemical potential is obtained from the continuity equation
\begin{align}
&0=-|e|A^{i,j}\frac{n_\uparrow^{i,j}+n_\downarrow^{i,j}}{\tau}=\Delta [\l^{i,j}(j_\uparrow^{i,j}+j_\downarrow^{i,j})]=\nonumber\\
&\sigma_\uparrow\Delta[\l^{i,j}(F^{i,j}-\delta\mu_\uparrow^{i,j})] +\sigma_\downarrow\Delta[\l^{i,j}(-F^{i,j} -\delta\mu_\downarrow^{i,j})]=\nonumber\\
&(\sigma_\uparrow+\sigma_\downarrow)\Delta\{\l^{i,j}[-\delta\mu^{i,j} +\P(F^{i,j}-\delta\mu_{\rm s}^{i,j})]\}\;.\nonumber\\\nonumber\\
&\rightarrow\Delta(\l^{i,j}\delta\mu^{i,j})=\P\Delta[\l^{i,j}( F^{i,j}-\delta\mu_{\rm s}^{i,j})]\;,
\end{align}
where the current polarization is given by
$\P=(\sigma_\uparrow-\sigma_\downarrow)/(\sigma_\uparrow+\sigma_\downarrow)$.
This result was already obtained for a continuous system in
Ref.~[\onlinecite{tserkovnyak2008}].
To find an equation for the spin accumulation, we write
\begin{align}\label{eq:appB}
&-|e|A^{i,j}\frac{n_\uparrow^{i,j}-n_\downarrow^{i,j}}{\tau} =\Delta[\l^{i,j}( j_\uparrow^{i,j}-j_\downarrow^{i,j})]=\nonumber\\
&\sigma_\uparrow\Delta[\l^{i,j}(F^{i,j}-\delta\mu_\uparrow^{i,j})] -\sigma_\downarrow\Delta[\l^{i,j}(-F^{i,j} -\delta\mu_\downarrow^{i,j})]=\nonumber\\
&(\sigma_\uparrow+\sigma_\downarrow)\Delta\{\l^{i,j}[F^{i,j} -\delta\mu_{\rm s}^{i,j}-\P\delta\mu^{i,j}]\}=\nonumber\\
&(\sigma_\uparrow+\sigma_\downarrow)(1-\P^2)\Delta[\l^{i,j}( F^{i,j}-\delta\mu_{\rm s}^{i,j})]\;.
\end{align}
If we compare this in the case of a square lattice to the expression in Ref.\onlinecite{tserkovnyak2008}
\begin{align}
\frac{1}{\lambda_\sd^2}\mu_\s-\nabla^2\mu_\s=-\nabvec\cdot\F\;,
\end{align}
we find that the density of spins that pile up can be expressed in terms of the spin accumulation as $(n_\uparrow^{i,j}-n_\downarrow^{i,j})/\tau=(\sigma_\uparrow+\sigma_{\downarrow}) (1-\P^2)\mu_{\rm s}^{i,j}/(|e|\lambda_{\rm sd}^2)$. We insert this expression to find that the spin accumulation on a lattice is determined by
\begin{align}
-\frac{1}{\lambda_{\rm sd}^2}\mu_{\rm s}^{i,j}=\frac{1}{A^{i,j}}\Delta[\l^{i,j}(F^{i,j}-\delta\mu_{\rm s}^{i,j})]\;.
\end{align}
\\

\end{document}